# Non-linear optics at twist interfaces in h-BN/SiC heterostructures

*Abhijit Biswas\*, Rui Xu, Gustavo A. Alvarez, Jin Zhang\*, Joyce Christiansen-Salameh, Anand B. Puthirath, Kory Burns, Jordan A. Hachtel, Tao Li, Sathvik Ajay Iyengar, Tia Gray, Chenxi Li, Xiang Zhang, Harikishan Kannan, Jacob Elkins, Tymofii S. Pieshkov, Robert Vajtai, A. Glen Birdwell, Mahesh R. Neupane, Elias J. Garratt, Tony G. Ivanov, Bradford B. Pate, Yuji Zhao, Hanyu Zhu\*, Zhiting Tian\*, Angel Rubio\*, and Pulickel M. Ajayan\**

Dr. Abhijit Biswas, Rui Xu, Dr. Anand B. Puthirath, Sathvik Ajay Iyengar, Tia Gray, Chenxi Li, Dr. Xiang Zhang, Dr. Harikishan Kannan, Jacob Elkins, Tymofii S. Pieshkov, Prof. Robert Vajtai, Prof. Hanyu Zhu, Prof. Pulickel M. Ajayan
Department of Materials Science and Nanoengineering, Rice University, Houston, Texas 77005, USA, Emails: **abhijit.biswas@rice.edu, hanyu.zhu@rice.edu, ajayan@rice.edu**

Gustavo A. Alvarez, Joyce Christiansen-Salameh, Prof. Zhiting Tian
Sibley School of Mechanical and Aerospace Engineering, Cornell University, Ithaca, NY 14853, USA, Email: **zhiting@cornell.edu**

Jin Zhang, Prof. Angel Rubio
Max Planck Institute for the Structure and Dynamics of Matter and Center for Free-Electron Laser Science, Luruper Chaussee 149, 22761 Germany, Email: **jin.zhang@mpsd.mpg.de**, **angel.rubio@mpsd.mpg.de**

Kory Burns
Department of Materials Science & Engineering, University of Virginia, Charlottesville, VA 22904, USA

Jordan A. Hachtel
Center for Nanophase Materials Sciences, Oak Ridge National Laboratory, Oak Ridge, TN 37831, USA

Tao Li, Prof. Yuji Zhao
Department of Electrical and Computer Engineering, Rice University, Houston, TX, 77005, USA

Dr. A. Glen Birdwell, Dr. Mahesh R. Neupane, Dr. Elias J. Garratt, Dr. Tony G. Ivanov
DEVCOM Army Research Laboratory, RF Devices and Circuits, Adelphi, Maryland 20783, USA

Dr. Bradford B. Pate
Chemistry Division, Naval Research Laboratory, Washington, DC, 20375 USA

Tymofii S. Pieshkov

Applied Physics Graduate Program, Smalley-Curl Institute, Rice University, Houston, TX, 77005, USA

Prof. Angel Rubio

Center for Computational Quantum Physics (CCQ), Flatiron Institute, New York 10010, New York, USA

Abhijit Biswas, Rui Xu, Gustavo A. Alvarez, Jin Zhang equally contributed to this work

**Table of Content**

We have grown h-BN thin films on SiC substrate exhibiting strong non-linear second-harmonic generation and ultra-low cross-plane thermal conductivity, attributed to the inherent formation of twisted nano-domain edges between the stacked h-BN nanocrystals with random in-plane orientations, as revealed by the first-principles time-dependent density functional theory.

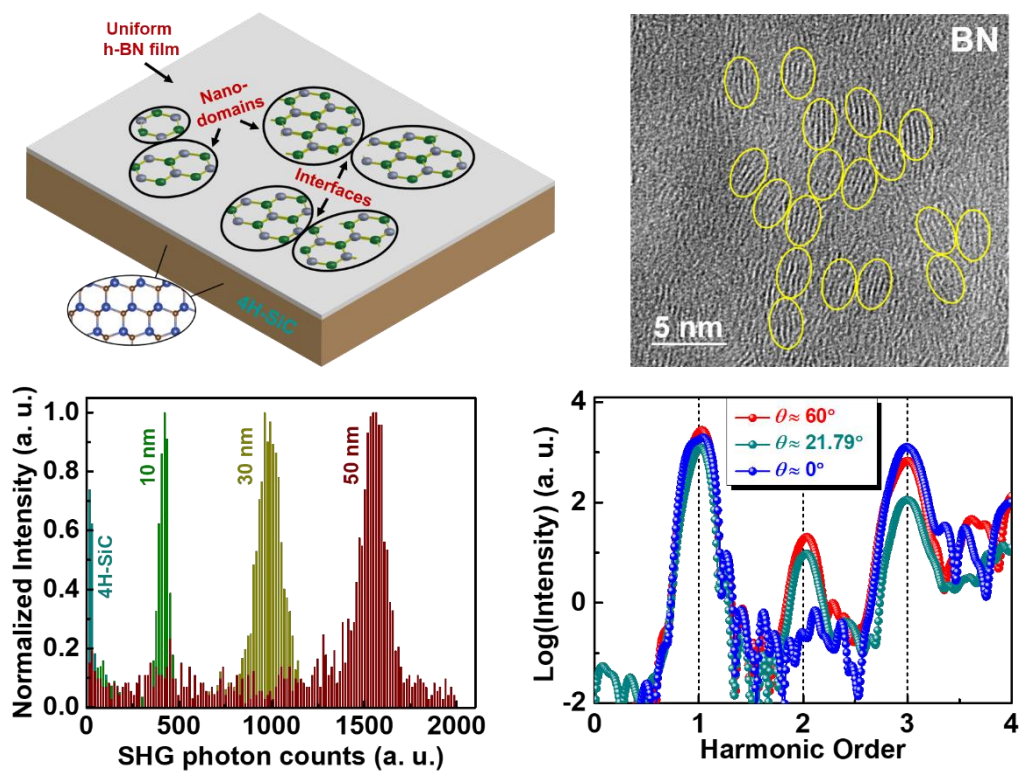

**Abstract**

Understanding the emergent electronic structure in twisted atomically thin layers has led to the exciting field of twistronics. However, practical applications of such systems are challenging since the specific angular correlations between the layers must be precisely controlled and the layers have to be single crystalline with uniform atomic ordering. Here, we suggest an alternative, simple and scalable approach where nanocrystalline two-dimensional (2D) film on three-dimensional (3D) substrates yield twisted-interface-dependent properties. Ultrawide-bandgap hexagonal boron nitride (h-BN) thin films are directly grown on high in-plane lattice mismatched wide-bandgap silicon carbide (4H-SiC) substrates to explore the twist-dependent structure-property correlations. Concurrently, nanocrystalline h-BN thin film shows strong non-linear second-harmonic generation and ultra-low cross-plane thermal conductivity at room temperature, which are attributed to the twisted domain edges between van der Waals stacked nanocrystals with random in-plane orientations. First-principles calculations based on time-dependent density functional theory manifest strong even-order optical nonlinearity in twisted h-BN layers. Our work unveils that directly deposited 2D nanocrystalline thin film on 3D substrates could provide easily accessible twist-interfaces, therefore enabling a simple and scalable approach to utilize the 2D-twistronics integrated in 3D material devices for next-generation nanotechnology.

Twisting of two-dimensional van der Waals (2D-vdW) materials and their heterostructures have recently become an emerging topic in materials science and condensed matter physics (coined as twistronics).[1-3] Twist angles between 2D-layers (via the formation of moiré patterns) alter electrical, optical and magnetic properties, manifesting a wide-range of multifunctional properties and twistronic systems.[4-6] Among numerous 2D materials, graphene and hexagonal boron nitride (h-BN) have been widely researched and their twisted h-BN/graphene/h-BN heterostructures (moiré superlattices) have been studied, demonstrating emergent phenomena.[7-10] Recently, studies have revealed "ferroelectric-like domains" due to the stacking of two h-BN layers at small twisted angles (θ <1°), attributed to the interfacial elastic deformations.[11-13] Ex-situ mechanical stacking of thin individual monolayers of exfoliated single crystal materials has been used for twistronics, which is non-trivial and a time-consuming process.[6] Although the transfer assembly enables successful stacking of 2D layers with desired twist angles, future application of twistronics would demand an alternative in-situ and clean approach for the mass production of twisted 2D-materials. Therefore, thin film growth of such 2D-materials with the inherent generation of twist-interfaces by exploiting the interfacial lattice mismatch between the films and underlying substrate templates would possibly be a feasible alternative method. Though recent studies on twisted epitaxial graphene on SiC have shown some promise as a platform for the fundamental study, the control of the twist angle between the layers remains technically challenging and non-trivial.[14-16]

Structurally, among various polymorphs of BN, the most stable h-BN lattice forms a 2D-layered structure with hexagonal stacking, with an ultrawide-bandgap (UWBG) of ~5.9 eV.[17,18] It is a centrosymmetric non-polar system with lattice parameters of $a = b = 2.50$ Å, $c = 6.661$ Å. It has unique functional properties, e.g., lightweight, chemical inertness, possesses excellent optical properties, anisotropic thermal conductivity, and high dielectric constant, making it an applicable inert and thin dielectric barrier as well as a heat-spreading/thermal isolation material in devices.[19] On the other hand, 4H-SiC is also a promising electronic material with a wide-bandgap (WBG) of ~3.2 eV.[17] It also forms a hexagonal lattice with a stacking sequence of ABCB and lattice parameters of $a = b = 3.08$ Å and $c = 10.08$ Å.[20] It has a unique combination of fascinating properties, e.g., high chemical stability, and high breakdown voltage, making it very useful for high-power field effect transistors, bipolar storage capacitors, and ultraviolet detectors.[21-23] Therefore, it could be envisioned that the growth of 2D h-BN on 3D 4H-SiC substrate with high in-plane lattice mismatch would require long-range commensurability, and to minimize (or release) the interfacial energy it may form unique structures with the consequent observation of emergent properties.

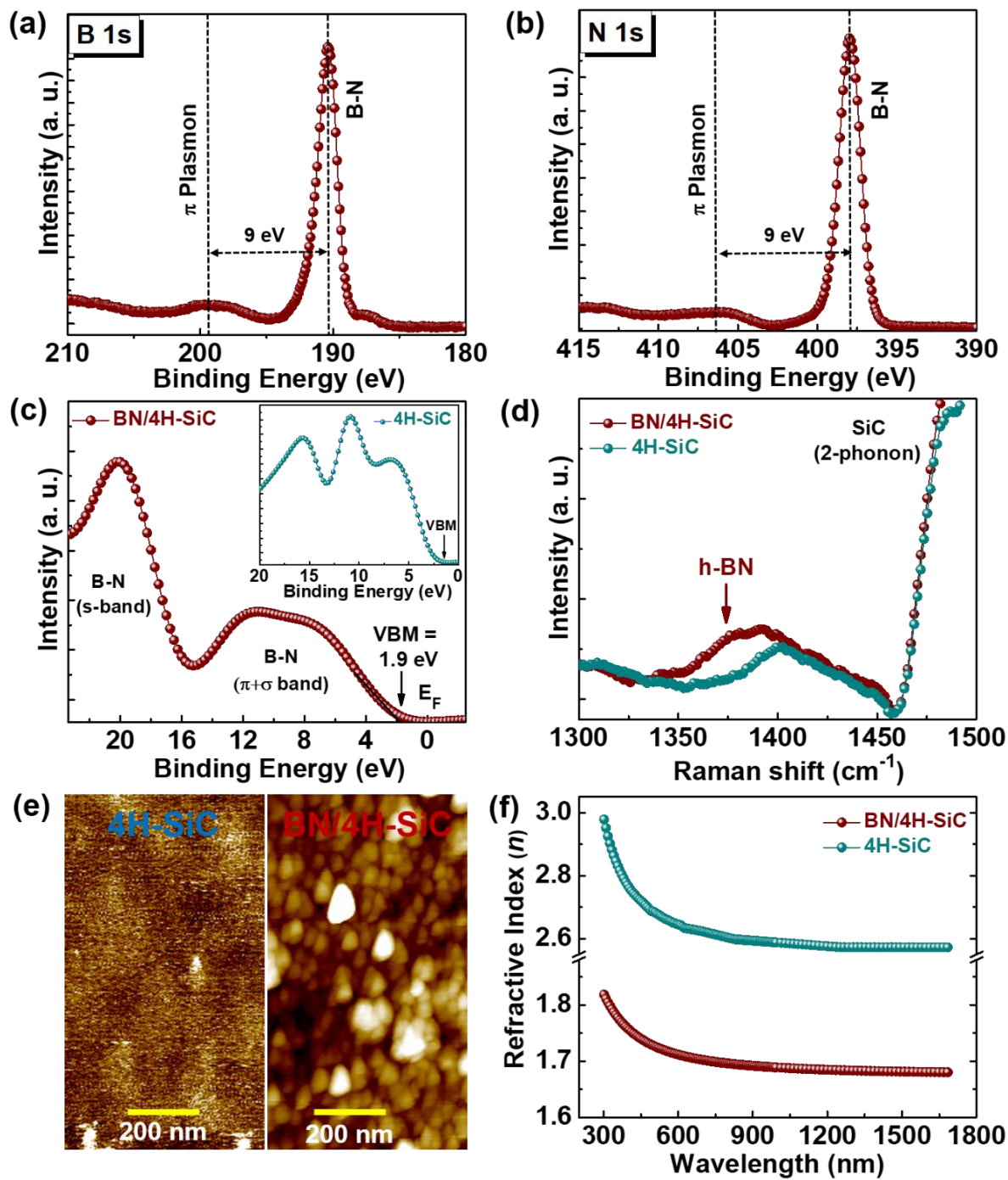

**Figure 1. Structural characterizations of h-BN films.** (a), (b) B 1s core and N 1s core X-ray photoelectron spectroscopy shows the characteristics of B-N bonding along with the π-Plasmon peaks. (c) XPS valence-band spectrum (VBS) of the grown h-BN film. Inset shows the VBS of 4H-SiC substrate. (d) Raman spectra show a hump within ~1360-1380 cm⁻¹, corresponding to the transverse optical E$_{2g}$ vibrations for in-plane B-N stretching in sp² bonded h-BN. (e) Atomic force microscopy surface morphology of 4H-SiC substrate and the BN film grown on it. (f) Refractive index of h-BN film and pristine 4H-SiC, in the visible to near infrared wavelength.

There have been several reports on h-BN thin film growth on pristine SiC or graphitized SiC substrates.[24-30] Recently, reports contradict the crystalline nature of the grown h-BN film on graphene covered 6H-SiC. For example, Shin *et al.* grew epitaxial single-crystal h-BN/graphene lateral structure on Si-terminated 4H/6H-SiC (0001).[29] Very recently, Lin *et al.* have shown by performing comprehensive analysis that on 6H-SiC (0001), h-BN forms a hexagonal $B_xN_y$ layer, rather than of high-quality stoichiometric epitaxial h-BN films.[30]

Herein, we have grown 2D h-BN thin films grown on 4H-SiC substrates and demonstrated its twist-dependent structure-property correlations. In-depth spectroscopic and microscopic characterizations confirm the growth of nanocrystalline h-BN film. The film is found to be strongly second harmonic generation (SHG) active, also demonstrating modulation of photon counts with film thickness. Simultaneously, we obtained low cross-plane thermal conductivity, at room temperature. These coexisting properties are attributed to the twisted interfaces related to nano-domain edges in h-BN with random in-plane orientations, as supported by the first-principles time-dependent density functional theory calculations.

**RESULTS AND DISCUSSION**

We have grown h-BN thin films on commercially available *n*-type 4H-SiC (0001) substrates by using the pulsed laser deposition (PLD) (**See experimental section**). We performed the core level X-ray photoelectron spectroscopy (XPS) to characterize the BN films. B 1s core and N 1s core spectra show the presence of characteristic B-N bonding (~190.3 eV and ~398.1 eV), along with π-Plasmon peaks that correspond to h-BN (**Figures 1a, 1b** and **S1**).[31-33] The atomic percentage of B:N was found to be ~1:1 (38.4:38.9 to be precise). The small hump around ~187.5 eV is due to the B-C peaks, possibly arising from an ambient air exposure related advantageous carbon effect. We also performed the XPS valence band spectroscopy (VBS) which shows the presence of two peaks, related to π+σ band and *s*-band (**Figure 1c**).[34,35] From VBS, we also found the valence band maxima (VBM) position at ~1.9 eV from the Fermi level ($E_F$). The characteristic VBS of 4H-SiC substrate is also shown (inset of **Figure 1c**).[36] Furthermore, the Raman spectra show the presence of $sp^2$ bonded $E_{2g}$ peak of h-BN around ~1370-1380 $cm^{-1}$, which corresponds to the in-plane $E_{2g}$ vibrational mode of h-BN (**Figure 1d**).[31] The BN film $E_{2g}$ peak intensity is significantly smaller than the surrounding two-phonon spectrum from the pristine 4H-SiC substrate. A similar trend in Raman spectra had also been observed in epitaxial graphene grown on SiC substrate.[37,38] and it is attributed to the strong interfering signal arising from the SiC. The Raman spectra at various locations on the film are also shown (**Figure S2**).

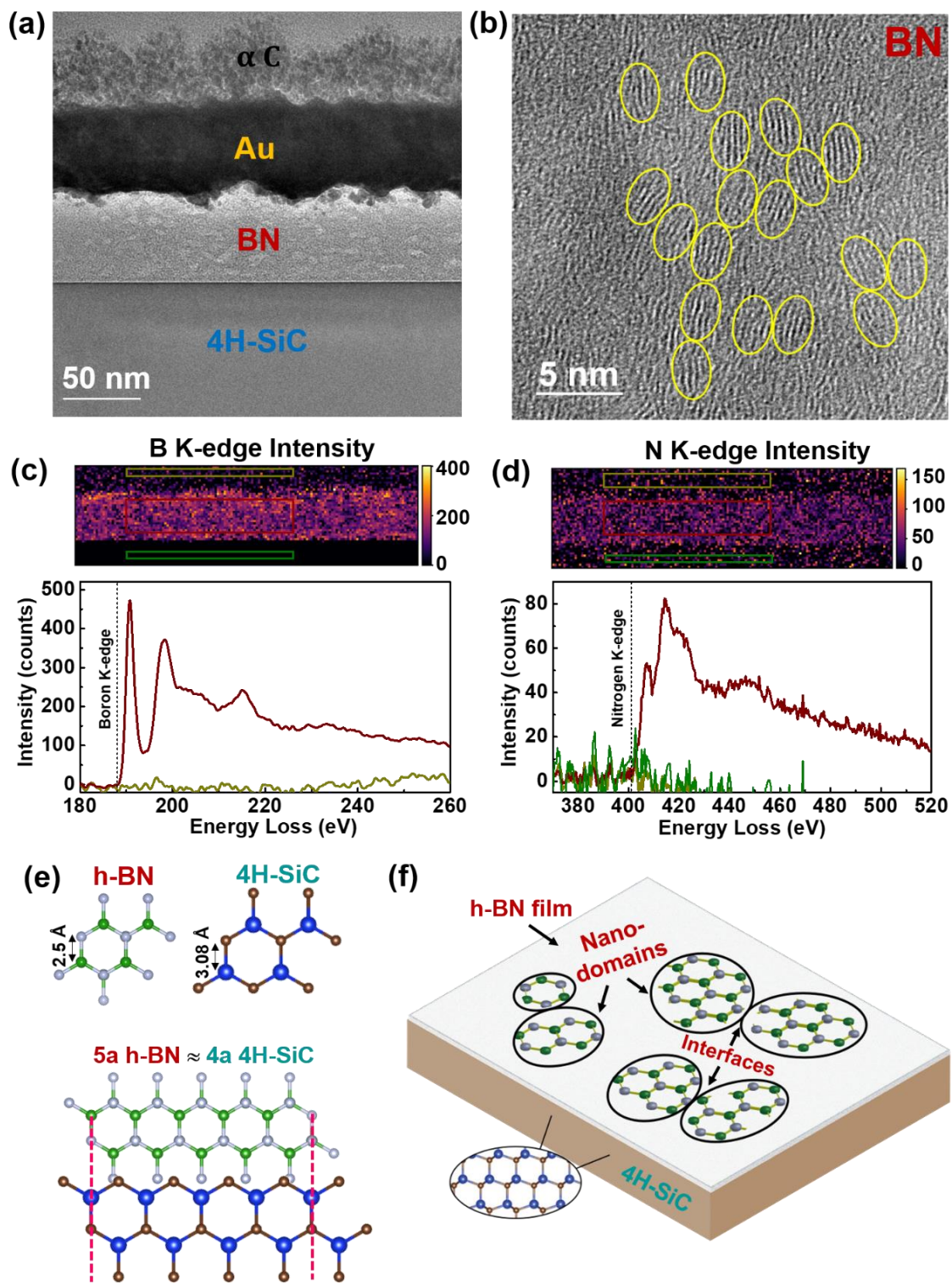

**Figure 2. Electron microscopy and crystal symmetry of h-BN/SiC.** (a) High-resolution cross-sectional microscopic imaging of h-BN film on SiC. (b) Several nano-domains with crystalline fringes are visible (yellow regions). (c), (d) Core-loss electron-energy loss spectra (EELS) and B K-edge and N K-edge elemental mapping showing the uniform presence of B and N. (e) Top-view in-plane hexagonal structures of h-BN and 4H-SiC. Commensurate lattice matching between h-BN and 4H-SiC. (f) Nano-domains of h-BN and possible twisted interfaces.

The atomic force microscopy (AFM) image shows the triangular-shape-like morphology with lateral sizes of ~50-100 nm (**Figure 1e**), compared with the smooth surface of bare single crystal 4H-SiC substrate. We also measured the refractive index (RI) of the film. In general, as shown in the literature, the RI value of crystalline h-BN flake is ~1.7-2.2, in the visible to near infrared (NIR) wavelength range, depending on the uncertainty of the thickness, roughness and crystallinity.[39,40] In our case, we also observed a comparable RI value of ~1.7-1.8, in the similar wavelength range (whereas 4H-SiC single crystal substrate shows RI of ~2.6-2.9) (**Figure 1f**). All these characterizations confirm the growth of h-BN films on 4H-SiC.

To get more information about crystalline quality of grown h-BN, we performed the cross-sectional high-resolution transmission electron microscopy (HRTEM) of h-BN/SiC film. For imaging, cross-sectional TEM lamella was prepared by a standard focused-ion-beam (FIB) method, based on mechanically thinning the sample followed by an Argon ion milling procedure. The cross-sectional bright-field image shows the interface between BN and SiC (**Figure 2a**). While zoomed into the BN regions, interestingly, nano-domains with clear fringes are visible, embedded in uniform amorphous-like BN (**Figure 2b**). The interplanar *d*-spacing of these fringes are ~0.33 nm, corresponds to the (0002) diffraction plane of h-BN (**Figure S3**).[18] Moreover, the electron-energy loss spectra (EELS) in the BN region also confirms that it is h-BN, with a uniform distribution of B and N (**Figures 2c** and **2d**). Regarding the nanocrystalline growth, it is quite plausible, as the lattice mismatch between the h-BN film and 4H-SiC substrate is extremely high, with a tensile strain of ~23.2% (calculated as $\left(\frac{a_s - a_f}{a_f}\right) \times 100\%$ where $a_s$ is the substrate and $a_f$ is the film in-plane lattice constant) (**Figure 2e**). Even the commensurate lattice matching (i.e. 5*a* h-BN ≈ 4*a* 4H-SiC) would impose a strain of ~1.44% (**Figure 2e**). Hence, to release such high interfacial energy, h-BN tends to form nanocrystalline ordering (**Figure 2f**), instead of epitaxial film with long-range ordered structure, thereby forming random nano-domains with grain boundaries and twisted-interfaces.

In addition to the h-BN/SiC interface, the twist angle generations within the 2D h-BN layers presents an intriguing phenomenon. It is possible that the short-range ordered random twisted nanodomains formed at the 2D/3D interface act as a template for propagating the twist within layers. With the growth process ongoing, the correlation or interaction between the layers has the potential to result in the spatial distribution of twisted nano-domains within the h-BN layers, extending throughout the bulk.[41,42] Scanning nanodiffraction 4D-STEM was also done to gain understanding of the crystalline structure (**Figure S4**). As seen, the film is amorphous-like, but present with crystallites in h-BN in the matrix, in which their positions were localized with random crystallographic orientations.

The non-linear optical properties of 2D-vdW materials are sensitive to the stacking order.[43,44] Broken inversion symmetry may lead to non-trivial topological electronic bands and more stable spin-orbit polarizations. In nanocrystalline h-BN, we expect various stacking faults resulting in local inversion symmetry breaking and cumulative finite SHG, which is prohibited in bulk single-crystalline h-BN. To test this hypothesis, we performed SHG imaging for h-BN films with various thicknesses grown on 4H-SiC. 4H-SiC belongs to space group P6₃mc and forbids SHG for incident light along [0001] direction. Meanwhile, single crystalline monolayer h-BN is known to have finite SHG with a second-order nonlinear susceptibility ($\chi^{(2)}$) about ~20 pm/V for monolayer h-BN.[45-47] For h-BN with an even number of layers, $\chi^{(2)}$ will vanish due to the restriction of structural symmetry.[45] However, for nanocrystalline h-BN with sub-wavelength domain sizes, defects and random domain orientation may statistically result in finite SHG regardless of layer number.[46,47]

**Figure S5** illustrates the schematics of our scanning SHG microscopy (**See experimental section**). While bare 4H-SiC substrate has almost zero SHG signal, h-BN/SiC shows strong SHG signal from the sample area with high contrast against the background (**Figures 3a** and **S6**). The histogram of SHG photon count distribution for BN/SiC with different film thickness are shown (**Figure 3b**). The effective $\chi^{(2)}$ of a 10 nm BN film is calculated to be ~3.3 pm/V, if we treat the film as an effective nonlinear crystal, based on the experimental parameters and calibration with known samples (**See experimental section**). For 50 nm BN film, the effective $\chi^{(2)}$ is ~1.2 pm/V. This reduction of effective $\chi^{(2)}$ in thicker films is expected as SHG from randomly oriented domains scales linearly with the thickness (**See experimental section**). We also measured the SHG for h-BN films grown on sapphire and GaN substrates (**Figure S7**), on which films are either single-crystalline (on sapphire) or fully disordered (on GaN).[48,49] Comparatively, the SHG signal for nanocrystalline h-BN on 4H-SiC is found to be significantly higher (~25 times) than the film on these substrates (with an effective $\chi^{(2)}$ of ~0.10 pm/V).[48]

We measured the cross-plane thermal conductivity $k_\perp$ of the film using the optical pump-probe method of frequency-domain thermoreflectance (FDTR) (**See experimental section**). An electro-optic modulator (EOM) induced a sinusoidal intensity modulation on the pump, 488 nm continuous wave laser (from a signal generated by the lock-in amplifier), creating a periodic heat flux on the sample surface.[50] An unmodulated, 532 nm continuous wave probe laser monitored the surface temperature through a change in surface reflectivity (**Figure S8**). Au was chosen as a transducer layer to maximize the coefficient of thermo-reflectance at the probe wavelength. We compared the measured phase lag of the probe beam (measured with respect to the reference signal from the lock-in amplifier) against the calculated phase lag of the sample surface temperature, induced by a periodic heat source at the sample surface.[51] The sample is modeled as a three-layer system, where each layer includes the volumetric heat capacity $c_p$, cross-plane thermal conductivity $k_\perp$, in-plane thermal conductivity $k_\parallel$, layer thickness, and the thermal boundary conductance $G_1$ and $G_2$ (inset of **Figure 3c**).

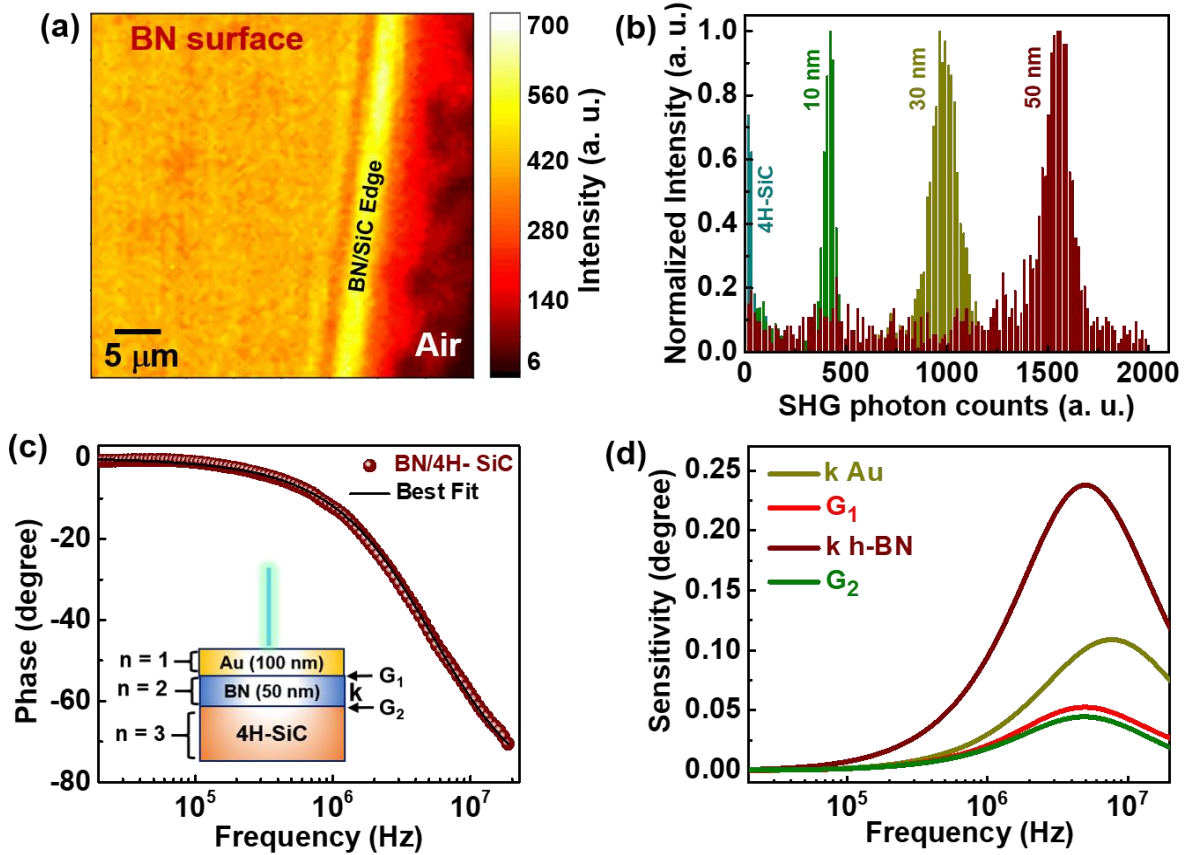

**Figure 3. Optical second harmonic generation and cross-plane thermal conductivity of h-BN.** (a) Spatial SHG intensity mapping of h-BN surface. (b) SHG photon count histogram of pristine 4H-SiC and h-BN/SiC with varying thickness. Large contrast can be observed from all three h-BN/SiC thin films compared with bare 4H-SiC. The histogram intensity is normalized by the highest count in each sample separately. (c) Phase vs. frequency data obtained from FDTR measurements shows a good approximation to the calculated best-fit curve. Inset shows the multilayer sample model. (d) Sensitivity analysis of the cross-plane thermal conductivity $k_\perp$ of the Au transducer layer, the thermal boundary conductance $G_1$, $G_2$, and the $k_\perp$ of h-BN.

An example of the phase vs frequency data obtained from FDTR of an average of three runs acquired on one spot location is shown (**Figure 3c**). The data is in good approximation to the best-fit curve obtained from solving the heat diffusion equation (**Figure S9**). A more comprehensive description of solving this equation is detailed by Schmidt et al.[51,52] We used the analytical method to estimate the uncertainty of our fitted data based on the film parameters and measurement.[52] Sensitivity analysis is used to help us determine which parameters can be fit together (**Supporting Information**). The cross-plane thermal conductivity $k_\perp$, of h-BN is the parameter of interest and it is most sensitive at higher frequencies (**Figure 3d**). The thermal boundary conductance $G_1$ between Au and the epitaxial h-BN, $G_2$ between h-BN and the bulk SiC substrate, and $k_\perp$ of Au are most sensitive at higher frequencies. Hence the focus was primarily on fitting the $k_\perp$ of h-BN. At room temperature, the average $k_\perp$ of three runs (measured at three different spot locations) of the grown h-BN film was found to be 0.47±0.04 $Wm^{-1}K^{-1}$, which is one order lower than the bulk crystalline h-BN.[53]

Based on the strong SHG signal and lower cross-plane thermal conductivity, we introduce the recently proposed concept of "twist-optics" by Yao et al., originating from the twisted interfaces.[54] They found that by tuning the twist-angles between the h-BN layers, SHG intensity could even be modulated by a factor of ~50 (for our nanocrystalline h-BN film it is ~25 times). Since the artificial alternation of layers can break the local symmetry at buried vdW-interfaces, the second harmonics waves generated at each interface can be coherently harnessed which amplifies the nonlinear efficiency and the observation of a strong SHG signal.

On the other hand, nanocrystalline h-BN shows lower cross-plane thermal conductivity than its single crystalline counterpart.[53] As reported in the literature, bulk h-BN shows anisotropy in thermal conductivity with room temperature cross-plane thermal conductivity varying between 2-5 $Wm^{-1}K^{-1}$.[53] Recently, Jaffe *et al.,* investigated grain boundary effect in the lowering of thermal conductivity as they introduced the twisted sheets of BN with different thickness and random twist angles.[55] They found the low cross-plane thermal conductivity of ~0.26±0.01 $Wm^{-1}K^{-1}$ for a ~74 nm film by stacking the five twisted h-BN layers, which is ~7 times lower than the calculated thermal conductivity value of ~2.0 $Wm^{-1}K^{-1}$. They attributed this lowering of cross-plane thermal conductivity to the twisted interfaces originating from the grain boundary effects. By using molecular-dynamics simulations it has been shown that twisted interfaces (i.e., grain boundaries) indeed act as strong phonon-phonon scattering centers, thus limiting the long phonon mean free paths at twist boundaries, and significantly reducing the cross-plane thermal conductivity.[56]

The situation of "twist-interfaces" and thereby the twist-optics scenario can be envisioned here by illustrating a correlation between the h-BN films structure and properties. Here h-BN film layers are not deliberately (mechanically) twisted on SiC substrate; however, as can be seen from the HRTEM image, the h-BN film layers have random nano-ordered regions, which can be imagined as twist interfaces. As a result, symmetry is broken at these interfaces with the generation of collective second harmonic signals, resulting in strong SHG. Simultaneously, since these interfaces are randomly ordered, they induce strong phonon-phonon interactions, with the reduction of phonon mean free paths, thereby lowering the thermal conductivity. Recently, it has been shown that the twist-angles between two SHG active hetero-bilayers of transition metal dichalcogenides (TMDCs) can also be calculated experimentally, where the twisted domain sizes are large enough (~few μm).[57,58] For our case, twisted domain are much smaller (~few nm), therefore remains difficult to resolve domains with specific twist angles.

For further insights about this twist-interface hypothesis, we performed extensive time-dependent density functional theory (TDDFT) calculations for higher harmonic generation (HHG) by twisting h-BN layers that have been demonstrated to exhibit a reliable description of the non-linear and non-perturbative responses of extended systems including h-BN heterostructures.[54, 59-61] Employing state-of-the-art simulations (**See experimental section**), we analyzed high harmonic spectra for three twist angles ($\theta = 0°$, 21.79°, and 60°) (**Figure 4a**). To reduce the complexity and computational burden, we have not incorporated the complicated stacking orders, strain effect and structural relaxation effects, rather focused on the role of the twist angle on the calculated HHG spectra.

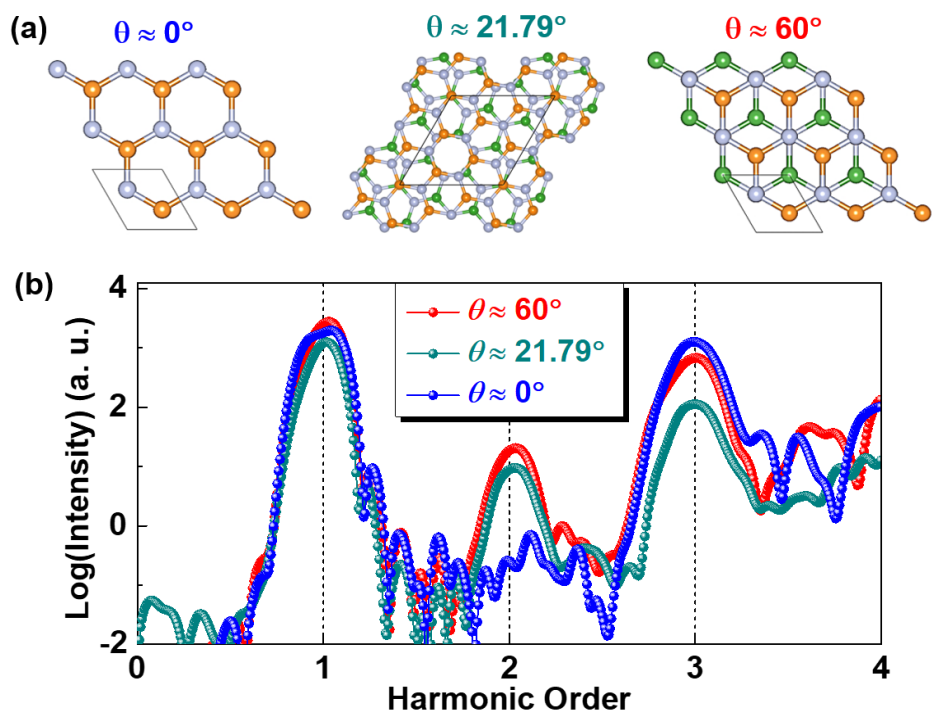

**Figure 4. Theoretical insights for the higher harmonic generation from twisted h-BN layers.** (a) Atomic structures of h-BN multilayers with different twist angles ($\theta$). The white, orange and green spheres are nitrogen, boron (top layers) and boron (bottom layers), respectively (b) Higher harmonic spectra for h-BN layers with a twist angle of ~0°, 21.79° and 60°, respectively, highlights the strength of second harmonics.

We find that the HHG spectrum of bulk h-BN changes significantly for the conditions with and without twists (**Figure 4b**). The harmonic intensity is probed along with the corresponding applied laser pulse, which is parallel with the pump laser polarization. For the pristine AA' stacking ($\theta = 0°$), clean odd harmonic peaks are observed for an excited pump laser (800 nm) while no clean even harmonics are obtained. However, for the $\theta = 21.79°$ and $60°$, HHG spectrum shows that the efficiency of second harmonics is significantly high. We observed ~27 and ~30 times increments in the second harmonic yield for the 21.79° and 60° twist cases, compared to the zero-twist angle (**Figure 4b**). These results are in good agreement with previous theoretical predictions.[62] The enhanced SHG is explained by the broken inversion symmetry from the phase difference in the adjacent layers in twisted h-BN layers, in contrast to the ideal non-twisted h-BN, which is an inversion-symmetric crystal in equilibrium.[54] Additionally, we carried out the corresponding simulations for the h-BN/SiC interface as well (**Figure S10**), which also shows the symmetry breaking and introduces significant yield of second harmonics, attributed to the h-BN/SiC interfacial electronic interactions.

In conclusion, h-BN thin films grown on 4H-SiC substrates exhibit fascinating structural-property correlation. Film shows strong second harmonic generation signal and ultra-low cross-plane thermal conductivity, similar to the mechanically twisted single crystalline h-BN layers. Concurrent observations of strong second harmonics generation and low cross-plane thermal conductivity are attributed to the twisted-interfaces (randomly oriented nanocrystalline domains), harnessing the second-harmonic signal, as well as increasing the phonon-phonon scattering (low thermal conductivity). First-principles time-dependent density functional theory calculations for twisted h-BN layers supports the observation of strong even-order harmonic signals. Our work demonstrates that directly deposited thin films of 2D-vdW materials on 3D substrates could have significant twisted interface regions and application worthiness, enabling simple, scalable approaches to utilize 2D twistronics integrated in 3D nanotechnology.

**Experimental Section**

*Thin film growth (Pulsed laser deposition, PLD):*

We have grown various thicknesses of h-BN thin films on commercially available *n*-type (Nitrogen doped) 4H-SiC (0001) substrates (MSE Suppliers, USA) by using the load-lock assisted PLD (operating with a KrF laser of 248 nm wavelength and a 25ns pulse width) growth facility. Prior to the insertion, we cleaned the substrate in an ultrasonic bath with acetone. The base pressure of the main growth chamber was ~5 ×10$^{-9}$ Torr, and the load-lock pressure was 5×10$^{-8}$ Torr. We used a commercially available one-inch diameter h-BN target for the ablation. Films are grown at ~750 °C, and under 100 mTorr N$_2$ partial pressure. Prior to the growth, substrates are pre-annealed at the same growth temperature and pressure for ~15 min. Films are grown at 5 Hz repetition rate and with the laser energy of ~230 mJ (fluency ~2.2 J/cm$^2$). The target-to-substrate distance was kept at ~50 mm. The laser spot size is ~1.5 mm × 7 mm. Films are post-annealed for ~15 min at the same growth pressure and temperature to compensate for the nitrogen vacancies, and then cooled down to room temperature at ~20 °C/min. The average growth rate is found to be ~3.6 nm/min. For experiments, we used ~3-5±0.02 mm$^2$ sizes films.

*Structural, chemical, microscopic and optical characterizations (XPS, VBS, Raman, AFM, HRTEM and Refractive index):*

X-ray photoelectron spectroscopy was performed by using PHI Quantera SXM scanning X-ray microprobe with 1486.6 eV monochromatic Al Kα X-ray source. High-resolution XPS elemental scans and valence band spectra (VBS) were recorded at 26 eV and 69 eV pass energy. Park NX20 AFM was used to obtain surface topography, operating in tapping mode using Al-coated Multi75Al cantilevers. Raman spectroscopy was measured using a Renishaw inVia confocal microscope with a 532 nm laser used as the excitation source. For cross-sectional atomic scale imaging, the TEM specimens were prepared via a focused ion beam (FIB) milling process employing a Helios NanoLab 660 FIB unit (at Rice University) with gold to avoid discharging and amorphous carbon as protecting layers. The EELS spectra and 4D-STEM image were acquired in Nion UltraSTEM unit (ORNL, USA) which was also equipped with a Gatan Enfina electron energy-loss (EEL) spectrometer to identify the elemental homogeneity of the sample with an EELS collection semi-angle of ~48 mrad. Variable angle spectroscopic ellipsometry (VASE) was applied to measure the refractive index (M-2000 Ellipsometer by J. A. Woollam Company) in the Class 100 (ISO 5) clean room facility at Rice University.

*Optical second harmonic generation (SHG):*

Second harmonic generation was performed using the home-built setup with a reflection geometry. A MaiTai (Spectra-Physics, USA) laser with 84 MHz repetition rate was used to give near-infrared pumping with the wavelength of 800nm and 40 fs pulse duration after optical compression. The laser is directed to a microscope with a scanning stage, which allows for spatially resolved SHG imaging, and then focused by a 20× objective to a spot size with a diameter of around 5.2 μm. The NA of the objective we used in the experiment is 0.45. The average laser power was kept at 10 mW at the sample location. The reflected signal is filtered by a 785 nm short-pass filter and a 400 nm bandpass filter to eliminate the reflected pump beam (**Figure S5**). The signal is finally detected using a single pixel photon counter (C11202-100, Hamamatsu, Japan). The calculation of effective $\chi^{(2)}$ is based on the method described elsewhere for ultra-thin non-linear optical materials,[63] followed by the equation:

$$P_{SHG} = \frac{16\sqrt{2}S\left|\chi^{(2)}_{eff}\right|^2 \omega^2 P_{in}^{\;2}}{c^3 \varepsilon_0 f \pi r^2 t (1+n)^6} \tag{1}$$

Where $P_{SHG}(P_{in})$ is the averaged SHG signal (pump) power, $\chi^2_{eff}$ is the effective $\chi^2$, S=0.94 is a shape factor for Gaussian pulses, $\omega$ is the frequency of pump light, $c$ is the speed of light, $\varepsilon_0$ is the vacuum dielectric constant, $f$ is the pump laser repetition rate, r is the radius of pump laser spot at focus, $t$ is the pulse duration, and $n$ is the refractive index of 4H-SiC substrate. We have also verified our calibration using monolayer CVD grown $WS_2$ samples with known $\chi^{(2)} \sim 0.7$ nm/V that after considering the collection efficiency for 400nm light (~10%), the output SHG power is about to be 4pW under the 1mW 800nm pumping pulse incidence.

The finite $\chi^{(2)}_{eff}$ of polycrystalline h-BN containing randomly oriented domains can be understood as follows. As mentioned in the main text, SHG signal will be perfectly cancelled in single-crystalline bulk h-BN or atomically thin h-BN containing even number of layers, where the second-order nonlinear susceptibility $\chi^{(2)}$ in adjacent layers have the same magnitude and exactly opposite directions. While in polycrystalline structures, a h-BN domain may have even or odd number of layers with equal probability, and the odd-numbered domain has finite SHG. With randomly orientated domains, the direction of $\chi^{(2)}$ from each domain are uncorrelated with each other, meaning the SHG emission from the sample is incoherent. In this case, SHG emission can be regarded as the sum of second harmonic dipole emission from different domains. When these domains are much smaller than the wavelength of SHG signal, and thus can be treated as point sources, the emission power collected by the objective lens, $P_{SHG}$, is given by [63]:

$$P_{SHG} = \frac{NA^2}{2} \frac{\mu_0 (2\omega)^4}{12\pi\epsilon_0 c^3} \left(\frac{2}{1+n_{sub}}\right)^2 |\sum_n p_n|^2 \tag{2}$$

Where NA=0.45 is the numerical aperture of the objective used in the experiment, $\epsilon_0$ is the vacuum dielectric constant, c is the speed of light, $p_n$ is the nonlinearly induced optical dipole moment of the $n^{th}$ domain within measurement region, and $n_{sub}$ is the refractive index of the substrate. The optical dipole moment induced by the incident pump electric field $E_\omega$ with a frequency of $\omega$, after considering the reflection from the substrate, can be expressed as:

$$|\sum_n p_n|^2 = \epsilon_0^2 \left(\frac{2}{1+n_{sub}}\right)^4 E_\omega^4 |\sum_n \chi_n^{(2)} V_n|^2 = \epsilon_0^2 \left(\frac{2}{1+n_{sub}}\right)^4 E_\omega^4 \left(\sum_n |\chi_n^{(2)} V_n|^2 + \sum_{m,n} \chi_m^{(2)} \chi_n^{(2)^*} V_m V_n\right) \tag{3}$$

Where $\chi_n^{(2)}$ and $V_n$ are the second order nonlinear optical susceptibility and volume of the $n^{th}$ domain, respectively. Since there is no correlation between the orientation of each domain, the expected value of the cross terms is zero. Thus, Eq (3) is simplified as:

$$P_{SHG} = \frac{NA^2}{2} \frac{\epsilon_0 (2\omega)^4}{12\pi c^3} \left(\frac{2}{1+n_{sub}}\right)^6 E_\omega^4 \sum_n |\chi_n^{(2)} V_n|^2 \tag{4}$$

Therefore, the total expected SHG for the randomly oriented h-BN nano domian is finite. Assume that the optical spot contains statistically many domains, we expect that $\sum_n |\chi_n^{(2)} V_n|^2 = \overline{\chi_n^{(2)}}^2 \overline{V_n}^2 n = \overline{\chi_n^{(2)}}^2 \frac{V^2}{n^2} n \propto \frac{V^2}{n}$, where V is the total volume of the sample illuminated by the pumping laser, and $n$ is the number of domains within the pumping laser spot. In summary, given the average domain size $\overline{V_n}^2$, the SHG from polycrystalline materials is proportional to the number of illuminated domains (i.e., thickness), and is roughly reduced by a factor of $n$ and distributed in all directions compared to single crystalline case where the emission is coherent and directional.

*Cross-plane thermal conductivity:*

We measured the cross-plane thermal conductivity $k_\perp$ by using the optical pump-probe method of frequency-domain thermoreflectance (FDTR).[64,65] An FDTR system is implemented with two continuous-wave lasers: a 488 nm pump and a 532 nm probe (**Figure S8**). The vertically polarized pump beam first travels through an optical isolator. The pump beam is focused into an electro-optic modulator (EOM), and a horizontally polarized beam is transmitted through a beam splitter (BS) and through a polarizing beam splitter (PBS). A microscope objective then focuses the beam onto the sample. The lock-in amplifier transmits a periodic signal to the EOM. The EOM then creates a periodic heat flux with a Gaussian spatial distribution on the sample surface. The probe beam first travels through an optical isolator and then through the BS, which coaxially aligns the probe beam with the pump beam. The probe beam, which is horizontally polarized, then travels through the PBS and passes through the quarter wave-plate, where the circularly polarized light is then focused by a microscope objective onto the sample (on the pump spot) to monitor the periodic fluctuations in reflectivity at the sample surface caused by the oscillating sample temperature. The post-sample is then reflected through the quarter-wave plate and reflected by the PBS to the photodetector.

***First-principles calculation:***

For calcualtions, we have chosen these specific values of twist angles ($\theta$ =0°, 21.79° and 60°) based on various practical (numerical) considerations (for arbitrary twist angles, much large supercells should be considered in real-time simulations, which are beyond our present computational capacity). This comparative analysis allows us to identify trends, patterns, or transitions that may exist across a range of values, providing a comprehensive understanding of the harmonic generations with different twisted angles. By studying the three twist angles, we can compare the properties or behaviors of different angles, adequate to the twist-angle dependence of second harmonic yields.

The non-linear response functions of the heterostructure were obtained by evaluating the time-dependent electronic current computed by propagating the Kohn-Sham equations in real space and real time, as implemented in the Octopus code.[66-68] with the adiabatic LDA functional.[69] All calculations were performed using the fully relativistic Hartwigsen, Goedecker, and Hutter (HGH) pseudopotentials.[70] In the simulations, we use three-layer h-BN on top and three-layer on the bottom to model the effect of interfaces. The laser pulses are treated classically in the dipole approximation (induced vector fields are imposed to be time-dependent but homogeneous in space) using the velocity gauge and we use a sin-square pulse envelope. The h-BN multilayers are sampled with six layers, including top three and bottom three layers, respectively. The real-space cell was sampled with a grid spacing of 0.4 Bohr and the Brillouin zone was sampled with a 42×42×21 k-point grid to sample the Brillouin zone, which yielded highly converged results for h-BN. The BN bond length is taken here as the experimental value of 1.445 Å. We consider a laser pulse of 25-fs duration at full-width half-maximum (FWHM) with a sin-square envelope, and the carrier wavelength λ is 800 nm, corresponding to 1.55 eV. The HHG spectra are directly calculated from the time-dependent current **J**(r,t) by a discrete Fourier transform after a temporal derivative as

$$HHG(\omega) = \left| FT\left(\frac{\partial}{\partial t}\int d^3r\, J(r,t)\right)\right|^2 \quad (5)$$

We did not considered a diagonal dielectric tensor or deal with all off-diagonal elements corresponding to anisotropy in high harmonic calculations in the dipole approximation. Neglecting the off-diagonal elements or assuming a diagonal dielectric tensor allows for a simpler model, which can be computationally more tractable and easier to interpret. In our case, the dielectric properties of the h-BN are not the primary focus, instead, the emphasis is on the interaction between the intense laser field and the electrons in the material. In addition, we mention that the computed dielectric response can be effectively treated as isotropic for in-plane polarization.

We have taken gradients along the zigzag direction of in-plane h-BN layers. We also checked the input laser polarization to perpendicular direction of in-plane h-BN layers. Our further calculations show the twist-angle dependence of second harmonics is robust (**Figure S11**).

For the h-BN/4H-SiC heterostructures HHG calculations, the supercell contains 164 atoms and the mismatch is 1.44% between 5×5 h-BN supercell and 4×4 4H-SiC slab, which is small enough to describe the experimental structure. To get the HHG spectrum of the heterostructure, we consider a laser pulse of 25-fs duration at full-width half-maximum with a sin-square envelope, and the carrier wavelength $\lambda$ is 800 nm, corresponding to 1.55 eV. This simulation takes about 24 hours with 5000 supercomputer cores.

**Data availability**

The data that support the findings of this study are available from the corresponding author upon reasonable request.

**Supporting Information**

Supporting information is available from the Wiley Online Library or the author

**Acknowledgments**

This work was sponsored partly by the Army Research Office and was accomplished under Cooperative Agreement Number W911NF-19-2-0269. The views and conclusions contained in this document are those of the authors and should not be interpreted as representing the official policies, either expressed or implied, of the Army Research Office or the U.S. Government. The U.S. Government is authorized to reproduce and distribute reprints for Government purposes notwithstanding any copyright notation herein. This work was sponsored by the Department of the Navy, Office of Naval Research under ONR Award no. N00014-22-1- 2357. R. X. and H. Z. are supported by the are supported by Welch Foundation C-2128. T. Li and Y. Zhao is supported as part of ULTRA, an Energy Frontier Research Center funded by the US Department of Energy (DOE), Office of Science, Basic Energy Sciences (BES), under Award No. DE-SC0021230. G.A.A. was sponsored by the National Science Foundation Graduate Research Fellowship under Grant No. 1650114 and by the GEM Associate Ph.D. Fellowship. This work is also supported by the Cluster of Excellence 'CUI: Advanced Imaging of Matter' of the Deutsche Forschungsgemeinschaft (DFG) - EXC 2056 - project ID 390715994, and SFB-925 "Light induced dynamics and control of correlated quantum systems" – project 170620586 of the Deutsche Forschungsgemeinschaft (DFG) and Grupos Consolidados (IT1453-22). We acknowledge support from the Max Planck-New York City Center for Non-Equilibrium Quantum Phenomena. This work was performed, in part, at the Cornell NanoScale Facility, a member of the National Nanotechnology Coordinated Infrastructure (NNCI), which was supported by the National Science Foundation (Grant No. NNCI2025233).

**Conflict of Interest**

The authors declare no conflict of interest.

**Data Availability Statement**

The data that support the findings of this study are available from the corresponding authors upon reasonable request.

Received: ((will be filled in by the editorial staff))

Revised: ((will be filled in by the editorial staff))

Published online: ((will be filled in by the editorial staff))

**Author Contributions**

A. B., R. V., and P. M. A. conceptualized the study. A.B., C. L., T. G., X. Z., S. A. I., H. K., T. P., and J. E., grew and characterized the films. A. B. P., K. B., and J. H. performed the FIB and electron microcopy. T. L. and Y. Z. performed the refractive index measurement. R. X., and H. Z. carried out the second harmonic optical measurement. G. A., J. C., and Z. T. measured the thermal conductivity. J. Z., and A. R. performed the theoretical calculations and analysis. A. G. B., M. R. N., E. J. G., B. B. P., and T. I. commented on the manuscript. All the authors discussed the results and contributed on the manuscript preparation.

**Supporting Information**

**Non-linear optics at twist interfaces in h-BN/SiC heterostructures**

*Abhijit Biswas\*, Rui Xu, Gustavo A. Alvarez, Jin Zhang\*, Joyce Christiansen-Salameh, Anand B. Puthirath, Kory Burns, Jordan A. Hachtel, Tao Li, Sathvik Ajay Iyengar, Tia Gray, Chenxi Li, Xiang Zhang, Harikishan Kannan, Jacob Elkins, Tymofii S. Pieshkov, Robert Vajtai, A. Glen Birdwell, Mahesh R. Neupane, Elias J. Garratt, Tony G. Ivanov, Bradford B. Pate, Yuji Zhao, Hanyu Zhu\*, Zhiting Tian\*, Angel Rubio\*, and Pulickel M. Ajayan\**

Abhijit Biswas, Rui Xu, Gustavo A. Alvarez, Jin Zhang equally contributed to this work

E-mails: **abhijit.biswas@rice.edu, jin.zhang@mpsd.mpg.de, hanyu.zhu@rice.edu, zhiting@cornell.edu, angel.rubio@mpsd.mpg.de, ajayan@rice.edu**

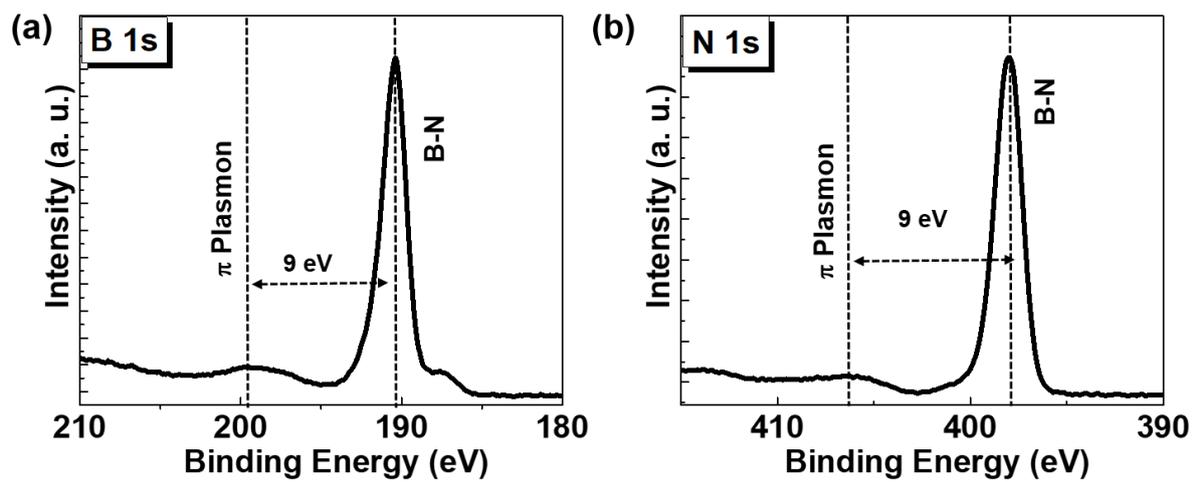

**Figure S1:** Structural characterizations of a 10 nm h-BN film. (a), (b) B 1s core and N 1s core X-ray photoelectron spectroscopy shows the characteristics of B-N peaks.

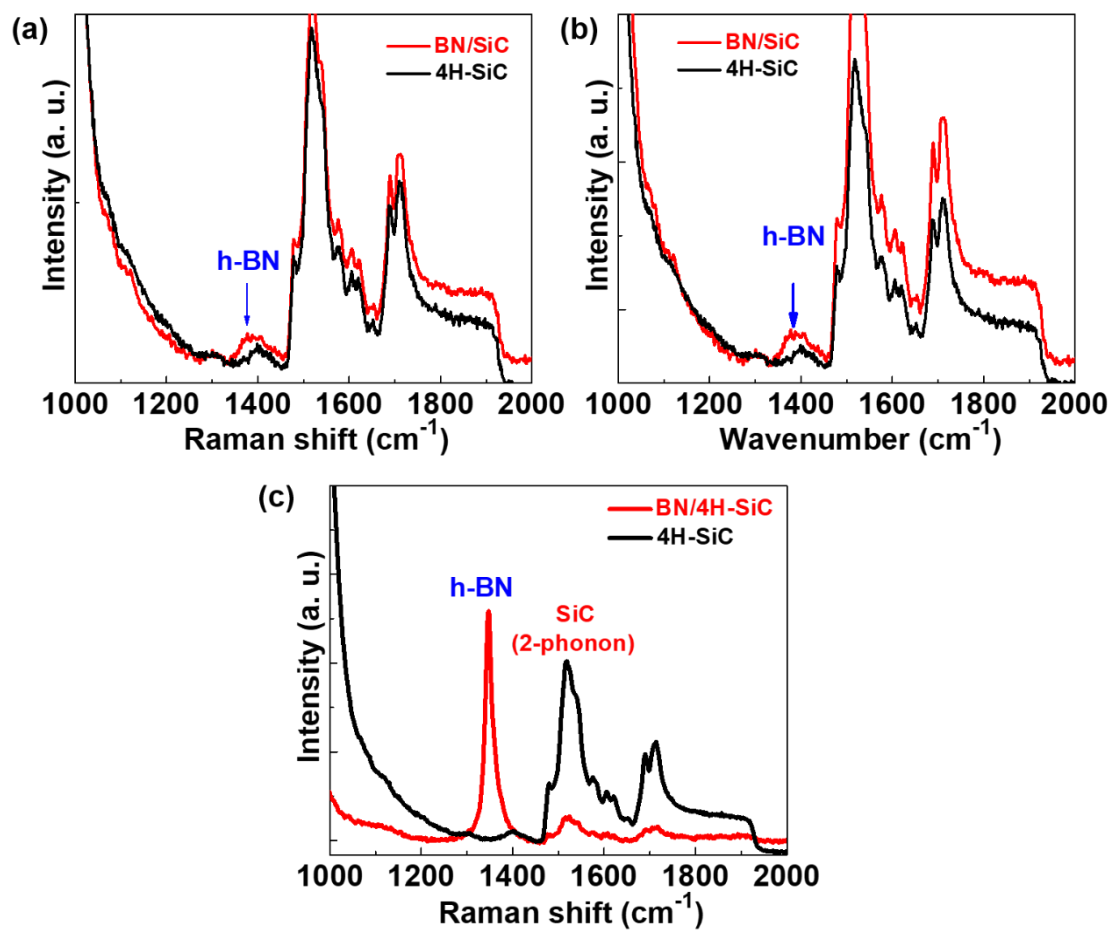

**Figure S2:** (a), (b) Raman spectra at several places of a thin h-BN film showing a hump around ~1360-1380 cm⁻¹, corresponding to the transverse optical E$_{2g}$ vibrations for in-plane B-N bond stretching in sp$^2$ bonded h-BN. (c) Raman spectra of a thick 300 nm BN/SiC film shows a clear E$_{2g}$ h-BN peak.

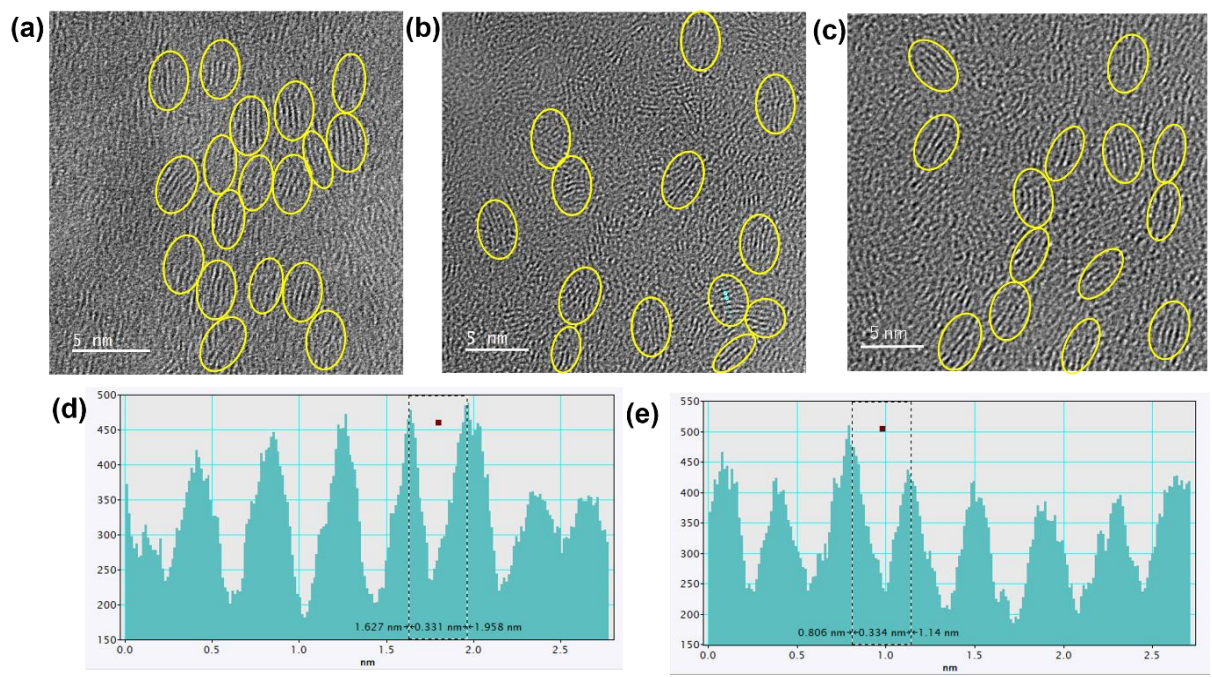

**Figure S3:** (a)-(e) Several nano-domain regions with clear crystalline fringes (*d*-spacing of ~0.33 nm) are also observed, corresponding to interplanar *d*-spacing of (0002) h-BN.

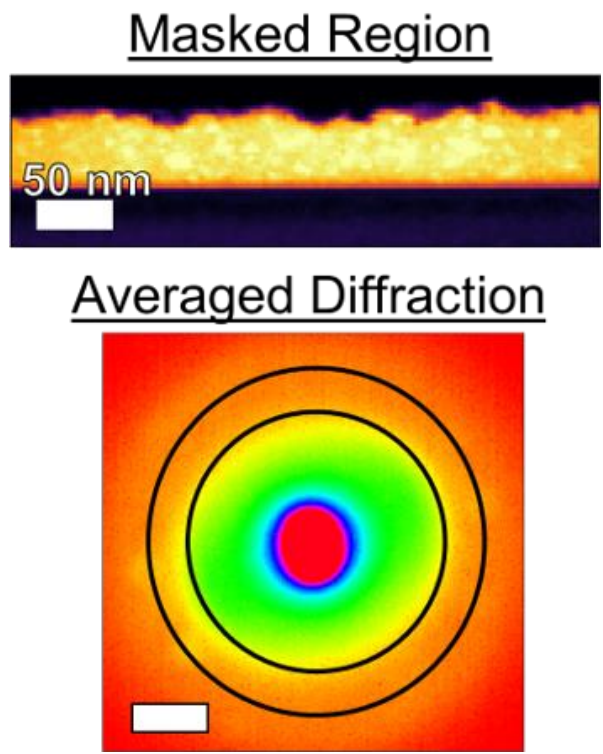

**Figure S4:** 4D-STEM analysis. The BN film is masked here so that the SiC and Au are omitted from visualizing the diffraction pattern. Mean diffraction from the different regions in the BN film. Weakly diffracted Bragg disks can be seen from the localized crystallites in the film. Scale bar is 5 mrad.

.

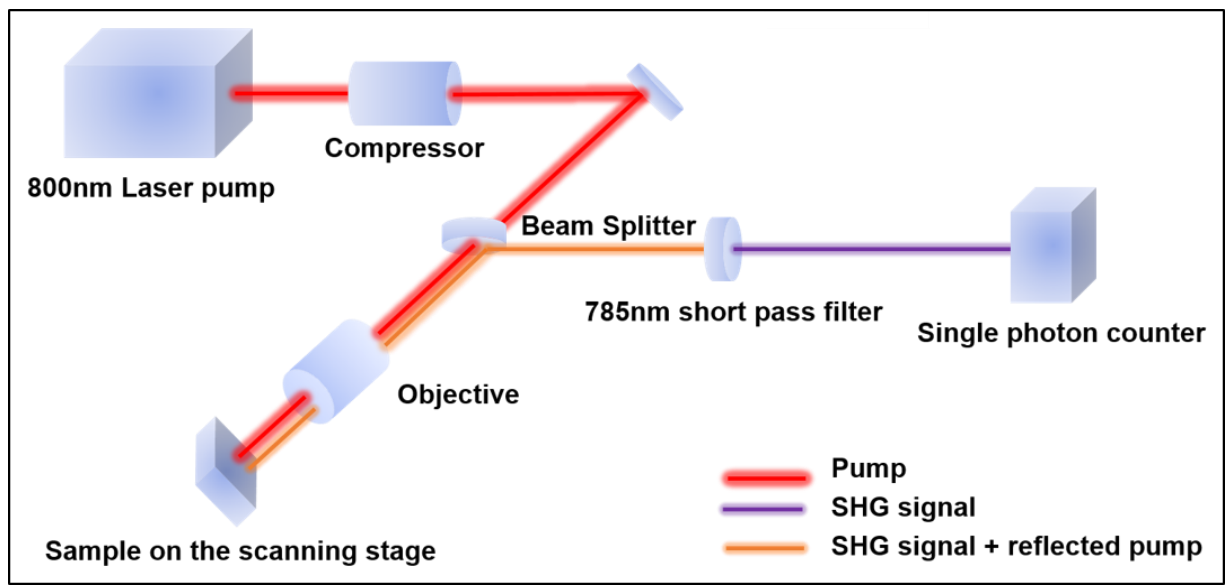

**Figure S5:** Schematic of the second harmonic generation (SHG) excitation and collection from the h-BN film surface.

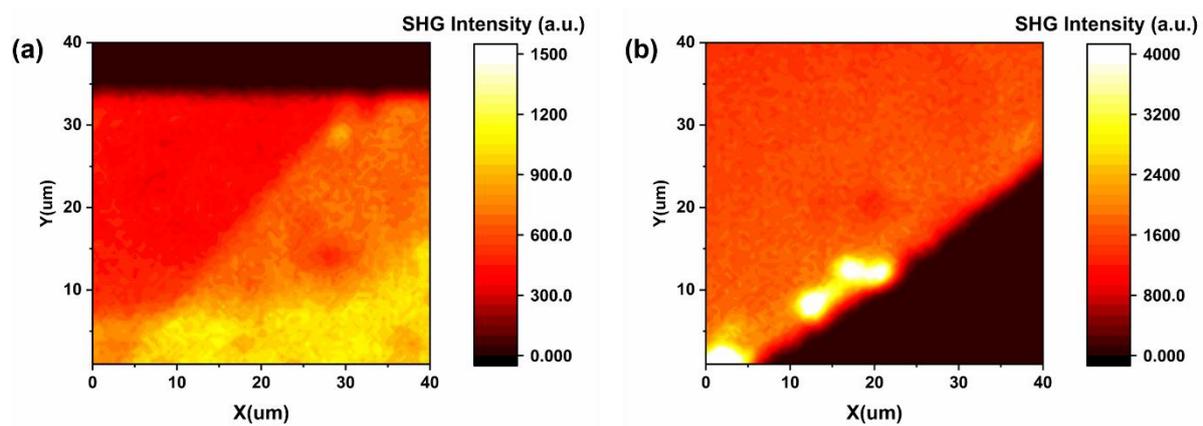

**Figure S6:** (a), (b) Spatial SHG intensity mapping of 30 nm (left) and 50 nm (right) h-BN films.

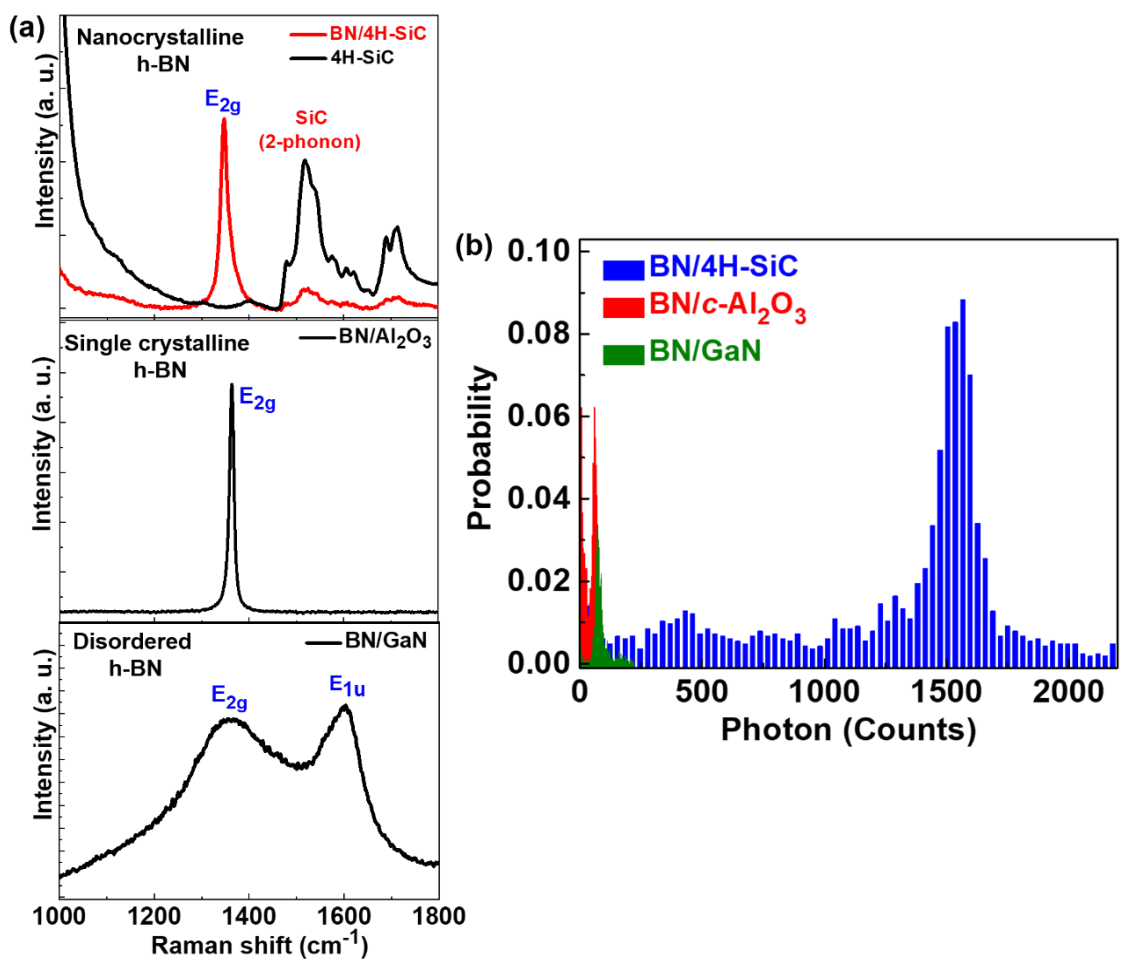

**Figure S7:** (a) Raman spectra, and (b) SHG response of h-BN thin films grown on SiC, *c*-Al₂O₃ and GaN substrates showing much stronger SHG signal for BN/SiC.

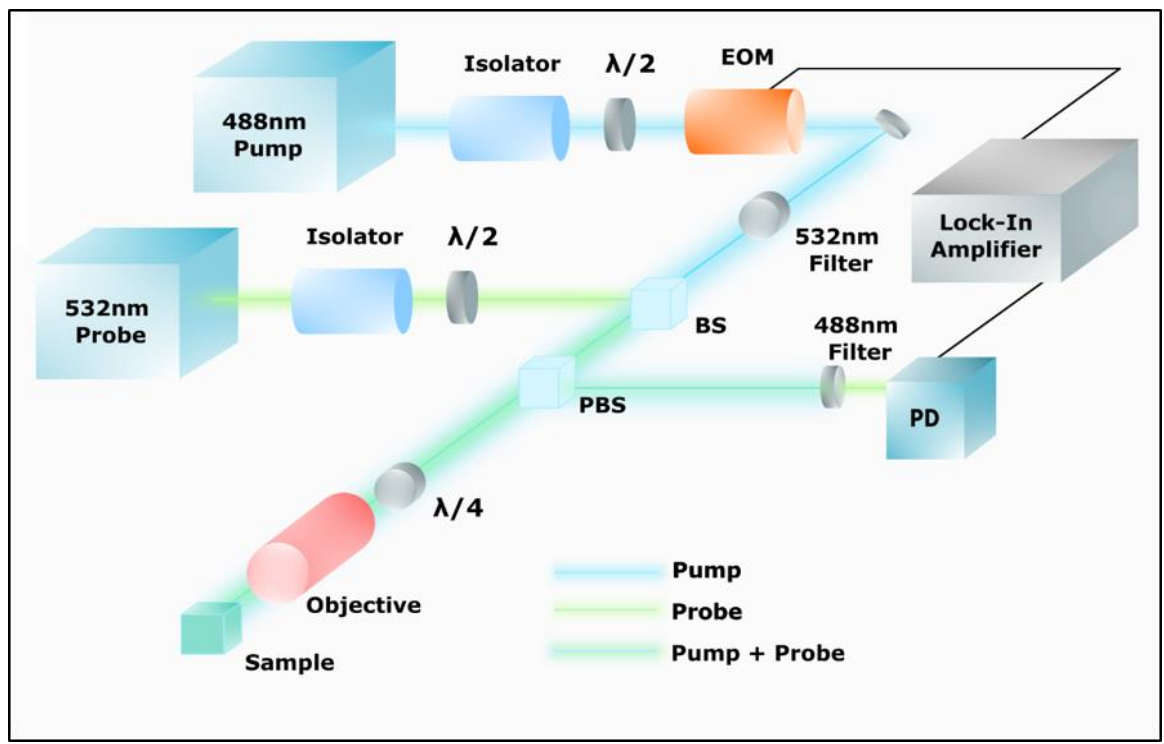

**Figure S8:** Schematic of the FDTR setup for the thermal conductivity measurement.

**Frequency-domain thermoreflectance (FDTR) details:**

The detailed derivation of the mathematical model can be found elsewhere.[51,71] The phase lag of the probe beam, measured with respect to the reference signal from the lock-in amplifier, is compared against the calculated phase lag of the sample surface temperature to a periodic Gaussian heat source at the sample surface.[51] Mathematically, the solution to the calculated phase lag (based on individual materials physical properties of interest, in our case $k$ and $G_2$) can be expressed as a complex number $Z(\omega_o)$, such that the output of the lock-in amplifier for a reference wave $e^{i\omega_o t}$ is given by

$$A e^{i(\omega_o t + \phi)} = Z(\omega_o) e^{i\omega_o t} \tag{S1}$$

where $\omega_o$ is the modulation frequency, $A$ is the amplitude, and $\phi$ the phase of the fundamental component of the probe signal with respect to the reference wave.[51]

In the case of continuous-wave pump and probe beams

$$Z(\omega_o) = \beta H(\omega_o) \tag{S2}$$

where $\beta$ is a factor including the thermoreflectance coefficient of the sample and the power of the pump and probe beams.[51] $H(\omega_o)$ is the thermal frequency response of the sample weighted by the intensity of the probe beam.[51] The weighted sample frequency response, $H(\omega_0)$, is obtained by solving the heat diffusion equation for a Gaussian heat source (the pump beam) impinging on a multilayer stack of materials and weighting the resulting temperature distribution at the top surface by the Gaussian intensity distribution of the probe beam.[51]

As an example, for a single slab of material in the frequency domain, the temperature $\theta_t$ and the heat flux $f_t$ on the top side of the slab are related to the temperature $\theta_b$ and the heat flux $f_b$ on the bottom side through

$$\begin{bmatrix} \theta_b \\ f_b \end{bmatrix} = \begin{bmatrix} \cosh(qd) & \dfrac{-1}{k_\perp q}\sinh(qd) \\ -k_\perp q \sinh(qd) & \cosh(qd) \end{bmatrix} \begin{bmatrix} \theta_t \\ f_t \end{bmatrix} \tag{S3}$$

where $d$ is the layer thickness, $k_\perp$ the cross-plane thermal conductivity and

$$q^2 = \frac{k_\parallel \mathcal{H}^2 + \rho ci\omega}{k_\perp} \tag{S4}$$

where $\mathcal{H}$ is the Hankel transfer variable, $\rho$ is the density, $c$ is the specific heat capacity, and $k_\parallel$ is the radial thermal conductivity.[2] The heat flux boundary condition at the top layer $f_t$ is given by the Hankel transform of a Gaussian spot with power $A_o$ and $1/e^2$ radius of the pump beam on the surface $w_0$

$$f_t = \frac{A_0}{2\pi} exp\left(\frac{-\mathcal{H}^2 w_0^2}{8}\right) \tag{S5}$$

Multiple layers are handled by multiplying the matrices for individual layers together

$$\begin{bmatrix} \theta_b \\ f_b \end{bmatrix} = \boldsymbol{M}_n \boldsymbol{M}_{n-1} \dots \boldsymbol{M}_1 = \begin{bmatrix} A & B \\ C & D \end{bmatrix}\begin{bmatrix} \theta_t \\ f_t \end{bmatrix} \tag{S6}$$

where $\boldsymbol{M}_n$ is the matrix for the bottom layer.[51] An interface conductance $G$ is treated by taking the limit as the heat capacity of a layer approaches zero and choosing $k_\perp$ and $d$ such that $G = k_\perp/d$ . Since we treat the nth layer as semi-infinite, Eq. (S5) reduces to

$$\theta_t = \frac{-D}{C} f_t \tag{S7}$$

The final frequency $H(\omega)$ in real space is found by taking the inverse Hankel of Eq. (S6) and weighting the results by the probe intensity distribution, which is taken as a Gaussian spot with $1/e^2$ radius of the probe beam on the surface $w_1$

$$H(\omega_0) = \frac{A_0}{2\pi} \int_0^\infty \mathcal{H}\left(\frac{-D}{C}\right) \exp\left[\frac{-\mathcal{H}^2(w_0^2 + w_1^2)}{8}\right] d\mathcal{H} \tag{S8}$$

This result is inserted into Eq. (S2), where the measurement of individual materials physical properties is performed as an inverse problem, minimizing the error between the lock-in phase data and the phase of Eq. (S2) via a non-linear least-squares algorithm.[51] Phase vs frequency data for each of the three spot measurments on the sample are shown below (**Figure S9**).

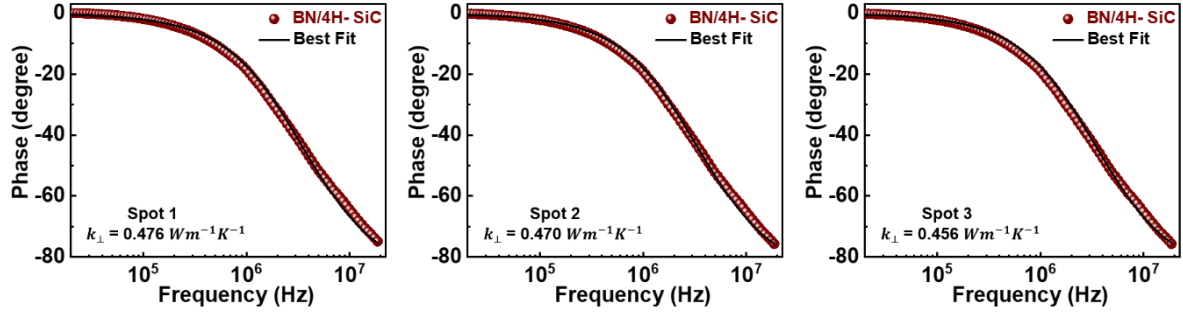

**Figure S9:** Phase vs. frequency data obtained from FDTR measurements shows a good approximation to the calculated best-fit curve. Each measurement is an average of three runs on three separate spots on the sample (i.e. nine measurements in total).

**Sensitivity Analysis:**

The sensitivity of the phase signal, $S_x$, to a parameter x is determined by

$$S_x = \frac{\partial \phi}{\partial \ln x} \tag{S9}$$

where $\phi$ is the phase in radians, expressed as

$$\phi = \tan^{-1} \frac{\Im(H(\omega))}{\Re(H(\omega))} \tag{S10}$$

where $\omega$ is the modulation frequency and H is the final frequency response of Eq. (S8). Eq. (S9) is evaluated for the thermal conductivity $k_\perp$ of the Au transducer layer, the thermal boundary conductance $G_1$ and $G_2$, and the $k_\perp$ of h-BN.

**Uncertainty Analysis:**

We used the analytical method to estimate the uncertainty of our fitted data based on the parameters and measurement.[52] We assumed uncertainty of 3% for the volumetric heat capacity,[52] and 4% for the thicknesses of the Au transducer layer and h-BN epilayer. To explore uncertainties of multiple unknown parameters, Jacobian matrices were used in the calculation. We selected the analytical method because it accumulates uncertainties from the parameters and measurements in the Jacobian matrices. The consideration of correlation would not overestimate the uncertainty.[72] Given the measured signal $\Phi$ and known parameter matrix $X_C$, the variance-covariance matrix of unknown matrix $X_U$ is

$$var(X_U) = (J_U' J_U)^{-1} J_U' (var(\Phi) + J_C var(X_C) J_C^{-1}) J_U (J_U' J_U)^{-1} \tag{S11}$$

Here, $var(\Phi)$ and $var(X_C)$ are the diagonal matrices whose elements are variances of measured signal and known parameters, respectively.[50] $J_C$ and $J_U$ are the Jacobian matrices of known and unknown parameters accordingly with the form:

$$J = \begin{pmatrix} \frac{\partial f(\omega_1, X)}{\partial x_1}|X^* & \cdots & \frac{\partial f(\omega_1, X)}{\partial x_N}|X^* \\ \vdots & \ddots & \vdots \\ \frac{\partial f(\omega_M, X)}{\partial x_1}|X^* & \cdots & \frac{\partial f(\omega_M, X)}{\partial x_N}|X^* \end{pmatrix} \tag{S12}$$

where $f(\omega_M, X)$ is the function to calculate the phase lag between pump and probe signals, $\omega_i$, $i = 1, \cdots, M$ are the frequency that takes the measurement, $x_j, j = 1, \cdots, N$ are the parameters and $X^*$ is the matrix of the fitted data.[52] The diagonal elements of $var(X_U)$ are the variances of the unknown parameters. Thus, this analysis consists of the propagation of errors and the variance among different measurement spots (i.e., the standard error of the six different spot locations, which is incorporated in $var(\Phi)$).

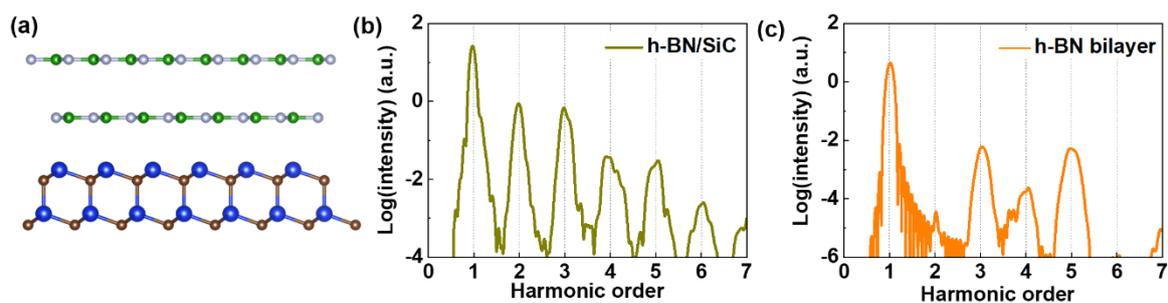

**Figure S10:** Theoretical insights for the high harmonic generation from twisted h-BN/SiC layers. (a) Atomic structures of the BN/SiC heterostructure. Top: h-BN bilayer with AA' stacking order and bottom: 4H-SiC slab. (b) High harmonic spectrum for the h-BN/SiC heterostructure. (c) High harmonic spectrum for h-BN bilayer.

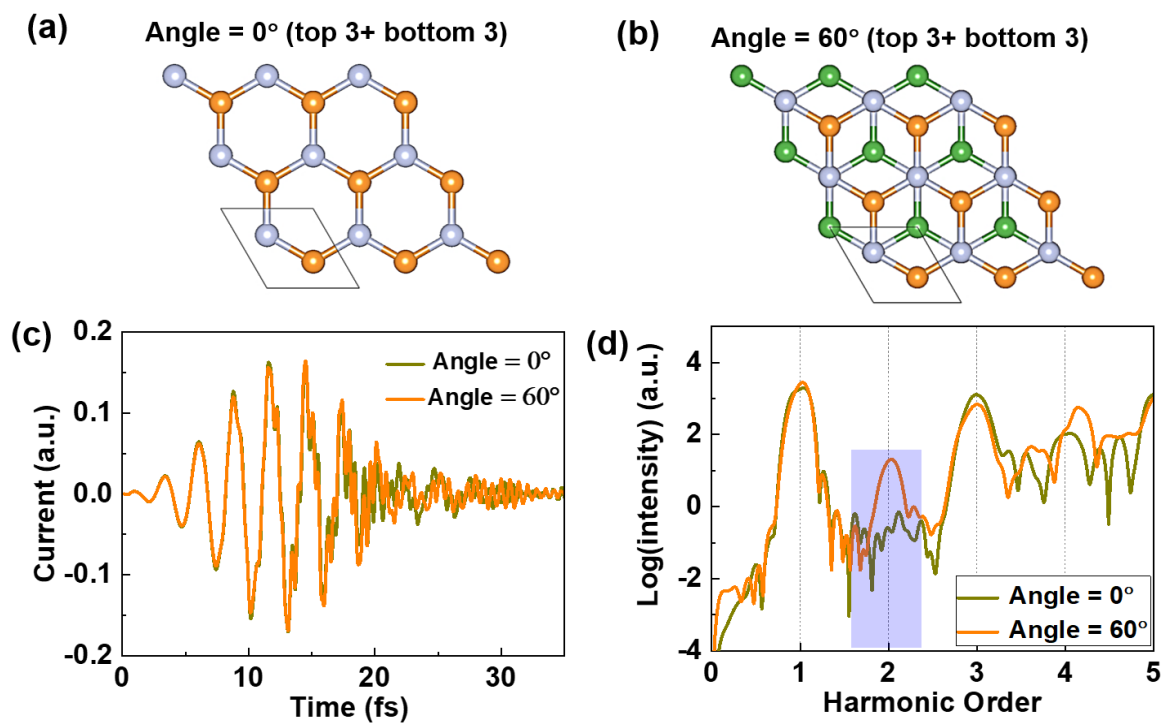

**Figure S11. Laser-induced total currents and corresponding high harmonic spectra in h-BN.** (a),(b) Atomic structures for the two-twisted angle. (c) Time-dependent electronic current for bulk h-BN, computed with different angles. (d) Higher harmonic spectra for h-BN layers with twist angles of 0° and 60°, respectively, highlights the strength of second harmonics. For the twist angle of 21.79°, the laser-induced total current is not applicable to compare directly with the twist angles of ~0° and 60° because the supercell changes the total numbers of atoms and electrons.